\documentclass[letterpaper, 10 pt, conference]{IEEEconf}
\IEEEoverridecommandlockouts
\usepackage{cite}
\usepackage{amsmath,amssymb,amsfonts}
\usepackage{algorithmic}
\usepackage{graphicx}
\usepackage{textcomp}
\usepackage{mathtools}
\usepackage{multirow}
\usepackage[table,xcdraw]{xcolor}
\usepackage{adjustbox}
\usepackage{booktabs}
\usepackage{breqn}

\def\BibTeX{{\rm B\kern-.05em{\sc i\kern-.025em b}\kern-.08em
    T\kern-.1667em\lower.7ex\hbox{E}\kern-.125emX}}

\title{\LARGE \bf Unsupervised Deformable Ultrasound Image Registration and Its Application for Vessel Segmentation

\thanks{$^{1}$Robotics Institute, Carnegie Mellon University, Pittsburgh, PA, USA {\tt\small (abhiman2, aorekhov, abal, choset, jgaleotti)@andrew.cmu.edu} }
}

\author{FNU Abhimanyu$^{1}$, Andrew L. Orekhov$^{1}$, Ananya Bal$^{1}$, John Galeotti$^{1}$, Howie Choset$^{1}$}

\begin{document}
\maketitle

\begin{abstract} 

This paper presents a deep-learning model for deformable registration of ultrasound images at online rates, which we call U-RAFT. As its name suggests, U-RAFT is based on RAFT, a convolutional neural network for estimating optical flow. U-RAFT, however, can be trained in an unsupervised manner and can generate synthetic images for training vessel segmentation models. We propose and compare the registration quality of different loss functions for training U-RAFT. We also show how our approach, together with a robot performing force-controlled scans, can be used to generate synthetic deformed images to significantly expand the size of a femoral vessel segmentation training dataset without the need for additional manual labeling. We validate our approach on both a silicone human tissue phantom as well as on \emph{in-vivo} porcine images. We show that U-RAFT generates synthetic ultrasound images with $98\%$ and $81\%$ structural similarity index measure (SSIM) to the real ultrasound images for the phantom and porcine datasets, respectively. We also demonstrate that synthetic deformed images from U-RAFT can be used as a data augmentation technique for vessel segmentation models to improve intersection-over-union (IoU) segmentation performance.
\end{abstract}

\section{Introduction} \label{sec:intro}

\par Vascular access and subsequent placement of central venous and arterial catheters is an essential first step for delivering life-saving medical care to trauma patients, e.g. administering anesthesia, monitoring vitals, and delivering rescuscitative treatments like Resuscitative Endovascular Balloon Occlusion of the Aorta (REBOA). Accessing a blood vessel, commonly done via the Seldinger technique \cite{seldinger2008catheter}, requires insertion of a needle into the center of the vessel, which is typically done by a highly skilled clinician using ultrasound to determine where to insert the needle. 
\par The work in this paper is motivated by the potential benefits of supporting human-guided vascular access with a robot so as to enable personnel away from centers of medical excellence to gain vascular access while avoiding vessel wall damage and hematomas caused by failed needle insertion attempts. This would be especially impactful on battlefields and in mass casualty disasters where there is limited access to trained medical personnel and hospital facilities. Examples of recent work towards the goal of robot-assisted femoral vessel access under ultrasound guidance include a hand-held device \cite{brattain2021femoral} and our group's system using a robot manipulator \cite{zevallos2021toward}.
\begin{figure}[ht]
  \centering
        \includegraphics[width=\columnwidth]{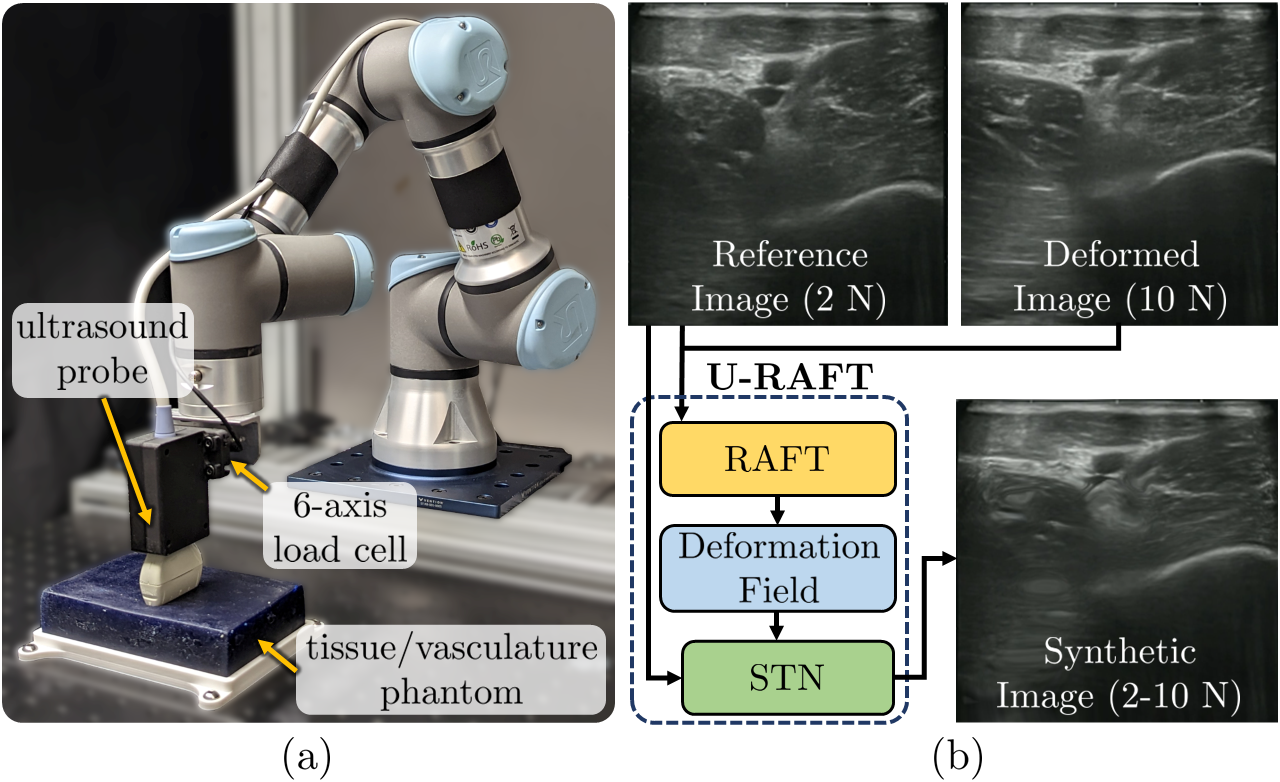}
    \caption{(a) The robot arm we used for capturing the ultrasound images with force-controlled scanning, together with a human tissue and vasculature phantom model we used to test U-RAFT. (b) Our U-RAFT model registers a deformed image to a reference image with RAFT, creating deformation field that is then used with a spatial transformer network (STN) to generate new deformed images, shown here with example \emph{in-vivo} porcine images.}
    \label{fig:system}
\end{figure}
\par Ultrasound imaging is an important modality for vascular access because it is safe, portable, and low-cost. However, the ultrasound probe needs to press against the skin to maintain acoustic coupling contact while capturing images, causing anatomical deformations during the scan. These deformations present a challenge for training deep neural networks that segment vessels from the ultrasound images.
\par Convolutional neural networks (CNNs), U-Net \cite{ronneberger2015u}, and variants of U-Net \cite{siddique2021unet_review} are commonly used in medical image segmentation, including vessel segmentation in ultrasound \cite{chen2021uncertainty, brattain2021femoral, marahrens2022towards, morales2023reslicing}. Training these models requires time-consuming labeling of the vessel contours in each image by personnel trained to interpret ultrasound images. Furthermore, the training set includes only a small subset of deformed vessel shapes that could occur, limiting the ability of the model to generalize to probe forces and deformed vessel shapes outside the training dataset. Although nonlinear warping augmentations could be applied to images \cite{ronneberger2015u}, this technique is not guaranteed to generate physically realistic image augmentations.
\par In this paper, we propose an approach to use deformable registration for augmenting images for training vessel segmentation models with small training datasets. Deformable registration is the problem of how to register pairs of images, one referred to as the \emph{reference image} and the other as the \emph{deformed image}, where the two images are of the same anatomy, but exhibit different deformations. By registering images captured at different forces, we will show that we can generate synthetic images at intermediate forces, ensuring that the augmented images are physically realistic. A similar idea was used in \cite{krivov2018mri,nalepa2019augmentation} for brain MRI images, and in \cite{tustison2019template} with lung MRI images.
\par Previous approaches to deformable registration include hand-crafted, iterative nonlinear optimization methods with a variety of cost function definitions and parameterizations of deformations \cite{che2017ultrasound,wang2018deformation_review,sotiras2013deformable_review}. These methods, however, typically have non-convex cost functions and are slow due to the large number of optimization parameters. To overcome these difficulties, deep-learning methods have also been presented \cite{boveiri2020review, fu2020review, zou2022review}, but work on deep-learning methods for ultrasound-to-ultrasound registration is limited. Compared to other higher-quality imaging modalities studied in most deformable registration work (e.g. CT-to-MRI or CT-to-ultrasound), ultrasound-to-ultrasound deformable registration is uniquely challenging due to noise, speckle, shadows, and mirror image artifacts \cite{che2017ultrasound}. Despite this, we will show that this problem is amenable to a deep-learning approach using our proposed model.
\par In this paper, we present a deep-learning model called U-RAFT (Unsupervised Recurrent All-pairs Field Transforms) for ultrasound-to-ultrasound deformable image registration and synthetic ultrasound image generation. As shown in Fig. \ref{fig:system}b, U-RAFT uses RAFT \cite{teed2020raft}, a CNN for optical flow estimation, to register images and create a deformation field (DF). It then uses a Spatial Transformer Network (STN) \cite{NIPS2015_33ceb07b} to generate new synthetic images. This approach allows for unsupervised training of U-RAFT, which we use to apply RAFT to ultrasound images for the first time, as well as generate realistic deformations for expanding vessel segmentation training datasets.
\par Compared to prior work, our work is unique in that it tackles ultrasound-to-ultrasound deformable registration in images of vasculature, we are able to register images at a rate suitable for online use ($\sim\,$33 Hz), and our training is unsupervised. We note that the utility of our deformable registration approach is not limited to vessel segmentation, since deformable registration in ultrasound images is broadly important for longitudinal studies/diagnosis, population studies, and intra-operative registration to anatomy \cite{che2017ultrasound,zou2022review}.
\par In Section \ref{sec:approach}, we present the network architecture of U-RAFT and describe three loss functions we considered for training this network in an unsupervised manner. In Section \ref{sec:results}, we present experimental results using U-RAFT on a benchtop 
silicone phantom model as well as \emph{in-vivo} porcine images of femoral arteries and veins. We compare the registration quality among the three loss functions we defined, and we demonstrate how the synthetic images that U-RAFT generates can be used as a data augmentation technique to improve the performance of a CNN for vessel segmentation. Section \ref{sec:conclusions} presents our conclusions and discussion on future directions.

\section{Approach}\label{sec:approach}

This section discusses the network architecture used for predicting the deformation field (DF) and the loss function used to train this network in an unsupervised manner. Furthermore, we discuss the use of the DF to generate new synthetic ultrasound images and their use to improve vessel segmentation.

\begin{figure*}[ht]
    \begin{center}
\includegraphics[width=\linewidth]{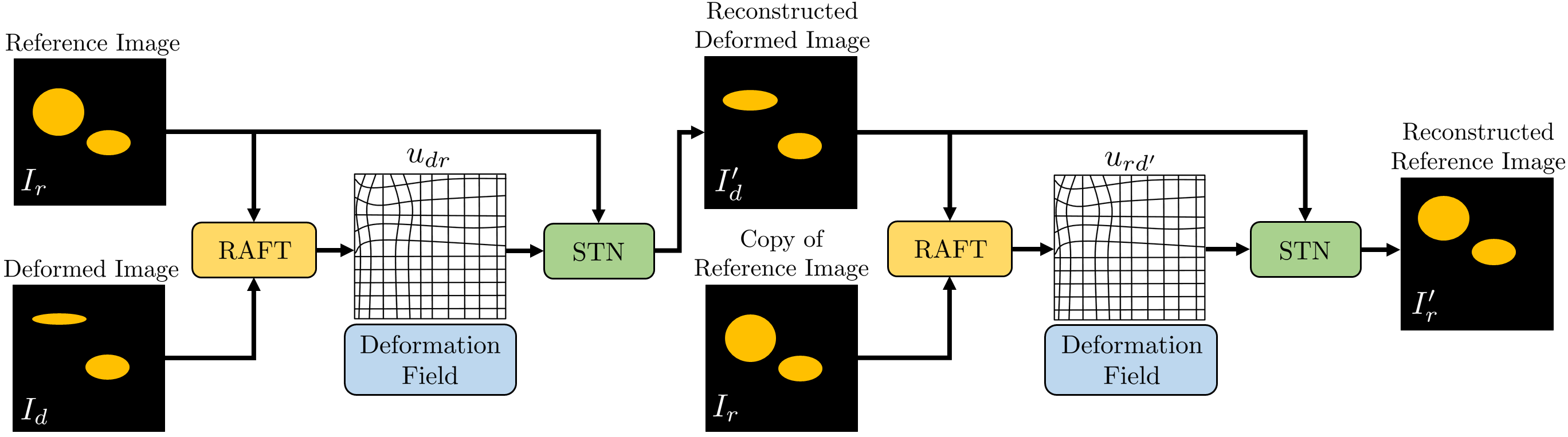}
    \end{center}
    \caption{Pipeline used to generate image reconstructions for training U-RAFT in an unsupervised manner. Our proposed cyclic loss function in \eqref{eq:cyclic_loss} improves registration quality by comparing the reconstructed reference image to the original reference image.}
    \label{fig:U-RAFT flowchart}
\end{figure*}

\subsection{U-RAFT network architecture}
Let $I_r$ and $I_d$ be the reference and deformed ultrasound images, respectively, collected at forces $F_r \in \mathbb{R}$ and $F_d\in \mathbb{R}$. We denote a DF as $u_{dr} = g_{\theta}(I_d, I_r)$, where $g_{\theta}$ is the function we seek to model with our network and the subscript $\theta$ denotes the network parameters used. Here, we use the state-of-art RAFT network\cite{teed2020raft} to model $g_{\theta}(I_d, I_r)$. We chose RAFT  over other CNN-based networks like FlowNet \cite{Dosovitskiy_2015_ICCV}, FlowNet2\cite{Ilg_2017_CVPR}, and PWC-Net \cite{Sun_2018_CVPR} because of its superior performance on the Sintel\cite{Mayer_2016_CVPR} and KITTI\cite{Menze2015CVPR} datasets. 
\par RAFT has been shown to outperform other optical flow methods in the RGB domain \cite{Dosovitskiy_2015_ICCV,Ilg_2017_CVPR,Sun_2018_CVPR}, but no prior work has shown the application of RAFT on medical ultrasound images, which are inherently noisier than RGB images\cite{che2017ultrasound}. RAFT is also a supervised method that needs a ground truth displacement field for training. Acquiring ground truth for ultrasound images is time-consuming and labor-intensive, so we seek to make the training unsupervised. We do this by passing the output of RAFT through a Spatial Transformer Network (STN) \cite{NIPS2015_33ceb07b} to generate a reconstructed deformed image $I_d' = \textrm{STN}(u_{dr}, I_{r})$.
This enables us to incorporate the similarity of $I_d'$ and $I_d$ in our training loss function, which, as we will show below, provides improved registration performance. We refer to the RAFT architecture together with STN as U-RAFT.
\subsection{Loss functions for unsupervised training}
We now define three different choices of loss functions to train the U-RAFT network in an unsupervised manner. We discuss the formulation and the advantages/disadvantages of each and perform a quantitative comparison between them in Sec. \ref{sec:results}-B.

The first loss function we consider, denoted as $\mathcal{L}_{us}$, consists of two parts $\mathcal{L}_{\textrm{ssim}}$, a multi-scale structural similarity (SSIM) loss that penalizes the differences in appearance between $I_d$ and $I'_d$, and $\mathcal{L}_{\textrm{smooth}}$, which penalizes abrupt changes in the neighboring pixels of $I'_d$ (generated from $u_{dr}$):
\begin{equation}
        \mathcal{L}_{\textrm{us}}(I_d,I_d',u_{dr}) = \beta \mathcal{L}_{\textrm{ssim}}(I_d,I_d') + (1-\beta)\mathcal{L}_{\textrm{smooth}} \left( u_{dr} \right)
        \label{unsupervisedRAFTLoss}
\end{equation}
where $\beta \in\mathbb{R}$ is a parameter to adjust the relative weight of $\mathcal{L}_{\textrm{ssim}}$ and $\mathcal{L}_{\textrm{smooth}}$. $\mathcal{L}_{\textrm{ssim}}$ and $\mathcal{L}_{\textrm{smooth}}$ are given by:
\begin{equation}
    \begin{split}
        \mathcal{L}_{\textrm{ssim}}(I_d,I_d') &= 1-\textrm{SSIM}(I_d,I_d') \\
        \mathcal{L}_{\textrm{smooth}}(u_{dr}) &= \textrm{mean}_{xy} \left( \frac{\nabla^2u_{dr}(x,y)}{\nabla x^2} + \frac{\nabla^2u_{dr}(x,y)}{\nabla y^2} \right)
        \label{eq:ssim}
    \end{split}
\end{equation}
where $x$, $y$ are the pixel location of a 2D-deformation field, and $\textrm{mean}_{xy}$ denotes the mean over all pixels.
\par The second loss function we consider is a cyclic version of \eqref{unsupervisedRAFTLoss}, denoted as as $\mathcal{L}_{\textrm{us-cyclic}}$. In this loss function, we register the reference image $I_r$ to the reconstructed deformed image $I_d'$ to generate a new reconstructed reference image $I_r'=\textrm{STN}(g_\theta(I_r,I_d'),I_d')$, as shown in Fig. \ref{fig:U-RAFT flowchart}. We then add to the loss function in \eqref{unsupervisedRAFTLoss} an additional term that calculates $\mathcal{L}_{\textrm{us}}$ for $I_r$, $I_r'$, and $u_{rd'}$.
\begin{equation}
         \mathcal{L}_{\textrm{us-cyclic}} =\mathcal{L}_{\textrm{us}}(I_{d},I_{d}',u_{dr}) + \mathcal{L}_{\textrm{us}}(I_{r},I_{r}',u_{rd'})
\label{eq:cyclic_loss}
\end{equation}

Finally, the third loss function we consider is designed to improve flow prediction in the vicinity of important anatomical features like veins, arteries, etc. We denote this feature-aware, cyclic, multi-scale SSIM loss function as $\mathcal{L}_{\textrm{fa-cyclic-us}}$. We use the scale-invariant feature transform (SIFT) algorithm \cite{lowe2004distinctive}, as implemented in OpenCV, to extract keypoints in an ultrasound image and construct a binary feature map around those keypoints. We then multiply each image by its binary feature map to create $\widetilde{I}_d$, $\widetilde{I}'_d$, $\widetilde{I}_r$, $\widetilde{I}'_r$  for the deformed, reconstructed deformed, reference, and reconstructed reference images, respectively. We then calculate the loss using the cyclic loss in \eqref{eq:cyclic_loss} but with the images with features extracted:
\begin{equation}
\mathcal{L}_{\textrm{fa-cyclic-us}} =\mathcal{L}_{\textrm{us}}(\widetilde{I}_{d},\widetilde{I}_{d}',u_{dr}) + \mathcal{L}_{\textrm{us}}(\widetilde{I}_{r},\widetilde{I}_{r}',u_{rd'})
\end{equation}

\subsection{Experimental setup and training details}
\par To train U-RAFT, we have collected ultrasound images from two different subjects: a human tissue/vasculature gel phantom model (CAE Blue Phantom), which we refer to as the \textit{blue-gel dataset}  and two different live pigs, which we refer to as the \textit{live-pig dataset}. The IACUC-approved \emph{in-vivo} porcine experiments were done in a controlled lab setting under the supervision of clinicians. 
The data was collected using a robotic ultrasound system which includes a UR3e manipulator  (Universal Robots) with a Fukuda Denshi portable point-of-care ultrasound scanner (POCUS) using a 5-12 MHz 2D linear transducer and a six-axis force/torque sensor (ATI)  mounted on the end effector, as shown in Fig. \ref{fig:system}. The datasets were collected either in a ``scanning mode'', where the robot scanned between two pre-defined points on the surface of the subject with a hybrid force motion controller similar to the controller described in \cite{goel2022autonomous}, or in ``palpation mode'', where the robot was commanded with a sinusoidal force profile at a single point on the skin surface. The minimum and maximum force used for both the modes were $2$\,N and $10$\,N, respectively. 

\par Using the loss functions introduced in Section \ref{sec:approach}-B, we train the U-RAFT network separately for the blue-gel and live-pig datasets.

For all these datasets, the RAFT weights were initialized with pre-trained KITTI weights\cite{Menze2015CVPR} and were trained for 150 additional epochs. The implementation is highly parallelized and performs full-batch gradient descent using the Stochastic Gradient Descent~\cite{bottou2012stochastic} optimizer in the Pytorch Autograd library\cite{paszke2017automatic}, with a batch size of 12 with a learning rate of 0.0001.
\subsection{Synthetic data generation}
Now that we can use U-RAFT to predict the DF between deformed and reference ultrasound images, we can generate synthetic ultrasound images at additional probe force values. Suppose ultrasound images at two different force values $F_r$ and $F_d$ are $I_{F_r}$ and $I_{F_d}$. We use U-RAFT to find the DFs $u_{F_r,F_d}= g_{\theta}(I_{F_r},I_{F_r})$ and $u_{F_d,F_r}= g_{\theta}(I_{F_d},I_{F_r})$. Then we use linear interpolation to find the intermediate DF $u_{F_{\textrm{new}},F_r}=\alpha u_{F_r,F_r}+(1-\alpha)u_{F_d,F_r}$ for an intermediate force $F_{\textrm{new}}$, where $\alpha = (F_{\textrm{new}}-F_{\textrm{min}})/(F_{\textrm{max}}-F_{\textrm{min}})$, with $F_{\textrm{min}}$ and $F_{\textrm{max}}$ being the minimum and maximum forces between which we interpolate. We then pass $u_{F_{\textrm{new}},F_r}$ through $STN$ to generate $I_{F_{\textrm{new}}}=\textrm{STN}(u_{F_{\textrm{new}},F_1}, I_{F_1})$. 
We then use synthetic images to augment the vessel segmentation dataset. We will show in Section \ref{sec:results}-B that this data augmentation technique helps a U-Net vessel segmentation model generalize to different forces.  
\section{Analysis and Results} \label{sec:results}
\begin{figure*}[ht]
    \begin{center}
\includegraphics[width=0.85\linewidth]{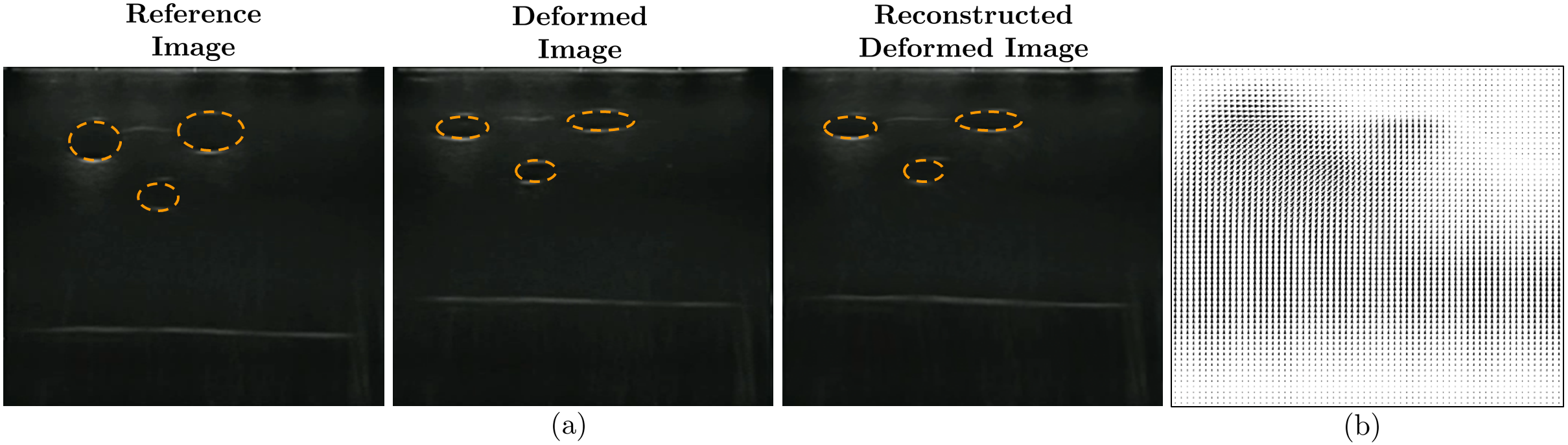}
    \end{center}
    \caption{Example deformable registration result from the blue-gel dataset, showing (a) the reference image $I_r$, the deformed image $I_d$, and the reconstructed deformed image $I_d'$, and (b) the deformation field calculated between $I_d$ and $I_r$ using U-RAFT, with a displacement vector plotted on a 4x4 pixel grid. The vessel walls for the blue-gel images are manually annotated for visibility.}
    \label{fig:uraft_reg_results}
\end{figure*}
\begin{figure}[h!]
    \begin{center}
        \includegraphics[width=1.0\linewidth]{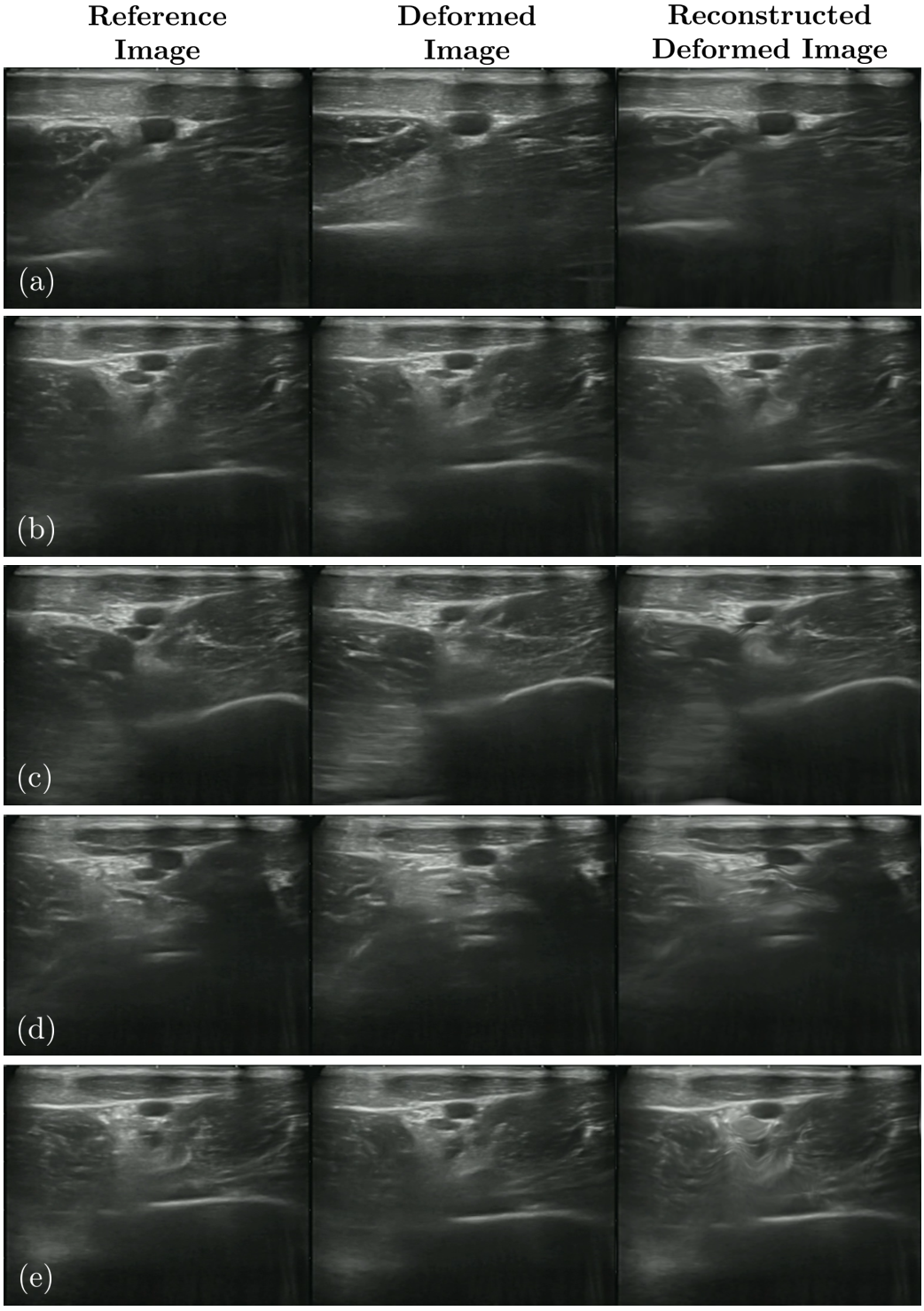}
    \end{center}
    \caption{Live-pig results showing reference, deformed, and reconstructed deformed images for (a) small deformation, (b) medium deformation, (c) large deformation, (d) vessel collapse, and (e) an atypical case of a vessel decollapsing. The similarity between the deformed and the reconstructed deformed images shows the efficacy of U-RAFT. Registration performance drops for the case of a vessel decollapsing, but even in this atypical scenario our approach fails gracefully.}
    \label{fig:pig_results}
\end{figure}
\subsection{Deformable registration results with U-RAFT}
In this section, we evaluate U-RAFT's performance on registering ultrasound images from the blue-gel and live-pig datasets using the three loss functions described in Section \ref{sec:approach}-B. We use the image similarity between the original deformed image $I_d$ and the reconstructed deformed image $I_d'$ to measure the efficacy of our method. Figure \ref{fig:uraft_reg_results} shows an example of  reference and deformed images from the blue-gel and the live-pig datasets along with the reconstructed reference images from the U-RAFT model. We use two metrics to compare the different loss functions: 1) SSIM\cite{sara2019image} and 2) a feature-aware-SSIM (F-SSIM), which is SSIM applied to the images after using SIFT to extract features as described in Section \ref{sec:approach}-B. 
\begin{figure}[t]
    \begin{center}
        \includegraphics[width=1.0\linewidth]{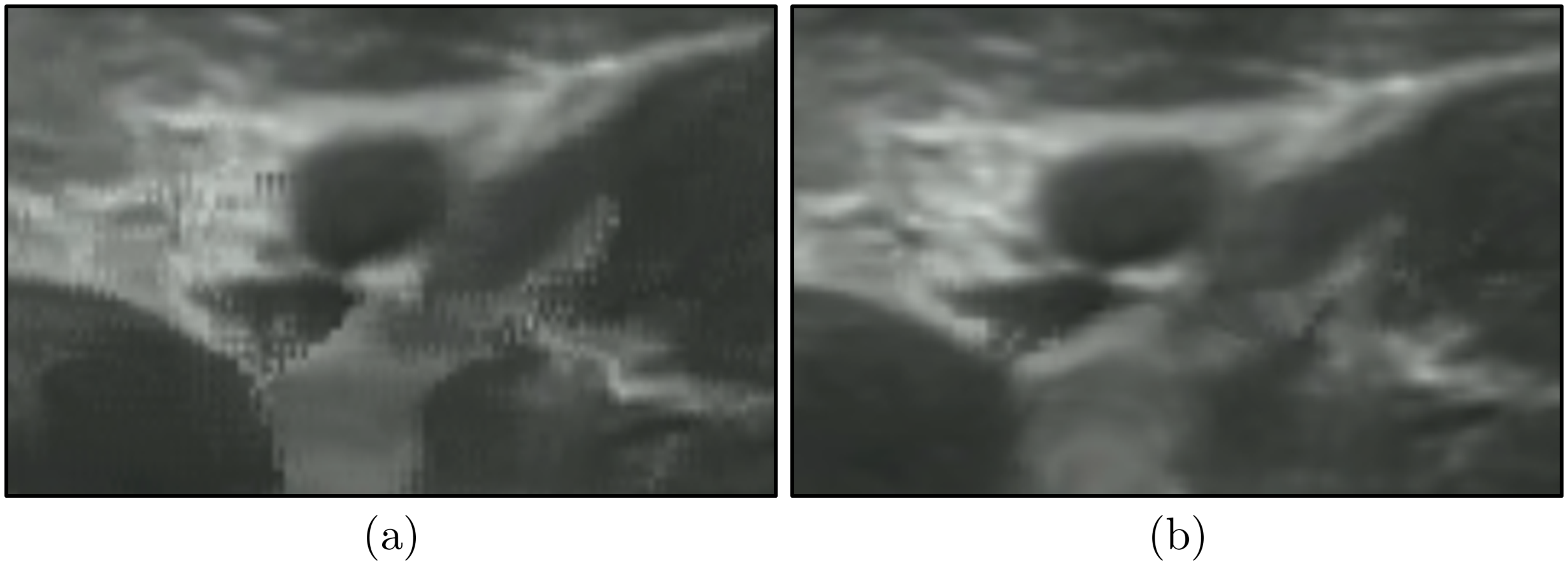} 
    \end{center}
    \caption{(a) Zoomed-in view of an example live-pig reconstructed deformed image $I_d'$ using $\mathcal{L}_{\textrm{us}}$. (b) Zoomed-in view of the reconstructed deformed image $I_d'$ using $\mathcal{L}_{\textrm{fa-cyclic-us}}$. The use of the feature-aware cyclic loss function helps remove the optical distortions observed in (a) for large deformations.} 
    \label{fig:ssim_vs_cyclic_ssim}
\end{figure}

The results are summarized in Table \ref{tab:loss_comparision}. In both SSIM and F-SSIM, the cyclic loss function $\mathcal{L}_{\textrm{cyclic-us}}$ outperforms $\mathcal{L}_{\textrm{us}}$, and the feature-aware cyclic loss function $\mathcal{L}_{\textrm{fa-cyclic-us}}$ outperforms the cyclic loss function $\mathcal{L}_{\textrm{cyclic-us}}$. The cyclic function outperforms the multi-scale structural similarity as it adds a better regularization of the flow prediction. We have also  observed qualitatively, as shown in the example in Fig. \ref{fig:ssim_vs_cyclic_ssim}, that the combination of feature extraction and cyclic loss leads to improved registration, particularly for larger deformations. 
%
%
\begin{table}[ht]
\begin{center}
\caption{Comparison of registration error for different loss functions. $\mathcal{L}_{\textrm{fa-cyclic-us}}$ outperforms the other two loss functions in terms of both SSIM and F-SSIM.}
\label{tab:loss_comparision}
\begin{tabular}{|c|c|c|c|c|}
\hline
\textbf{Loss Function} & \textbf{\begin{tabular}[c]{@{}c@{}}SSIM\\  (blue-gel)\end{tabular}} & \textbf{\begin{tabular}[c]{@{}c@{}}F-SSIM\\ (blue-gel)\end{tabular}} & \textbf{\begin{tabular}[c]{@{}c@{}}SSIM\\ (live-pig)\end{tabular}} & \textbf{\begin{tabular}[c]{@{}c@{}}F-SSIM \\ (live-pig)\end{tabular}} \\ \hline
$\mathcal{L}_{\textrm{us}}$ & 0.905 & 0.966 &0.870  & 0.918 \\ \hline
\begin{tabular}[c]{@{}c@{}}$\mathcal{L}_{\textrm{cyclic-us}}$\end{tabular} &0.907  &0.967  &0.883  &0.927  \\ \hline
\begin{tabular}[c]{@{}c@{}}$\mathcal{L}_{\textrm{fa-cyclic-us}}$\end{tabular} &0.909  &0.969  &0.886  &0.931  \\ \hline
\end{tabular}
\end{center}
\end{table}

\begin{figure*}[ht]
    \begin{center}
        \includegraphics[width=0.9\linewidth]{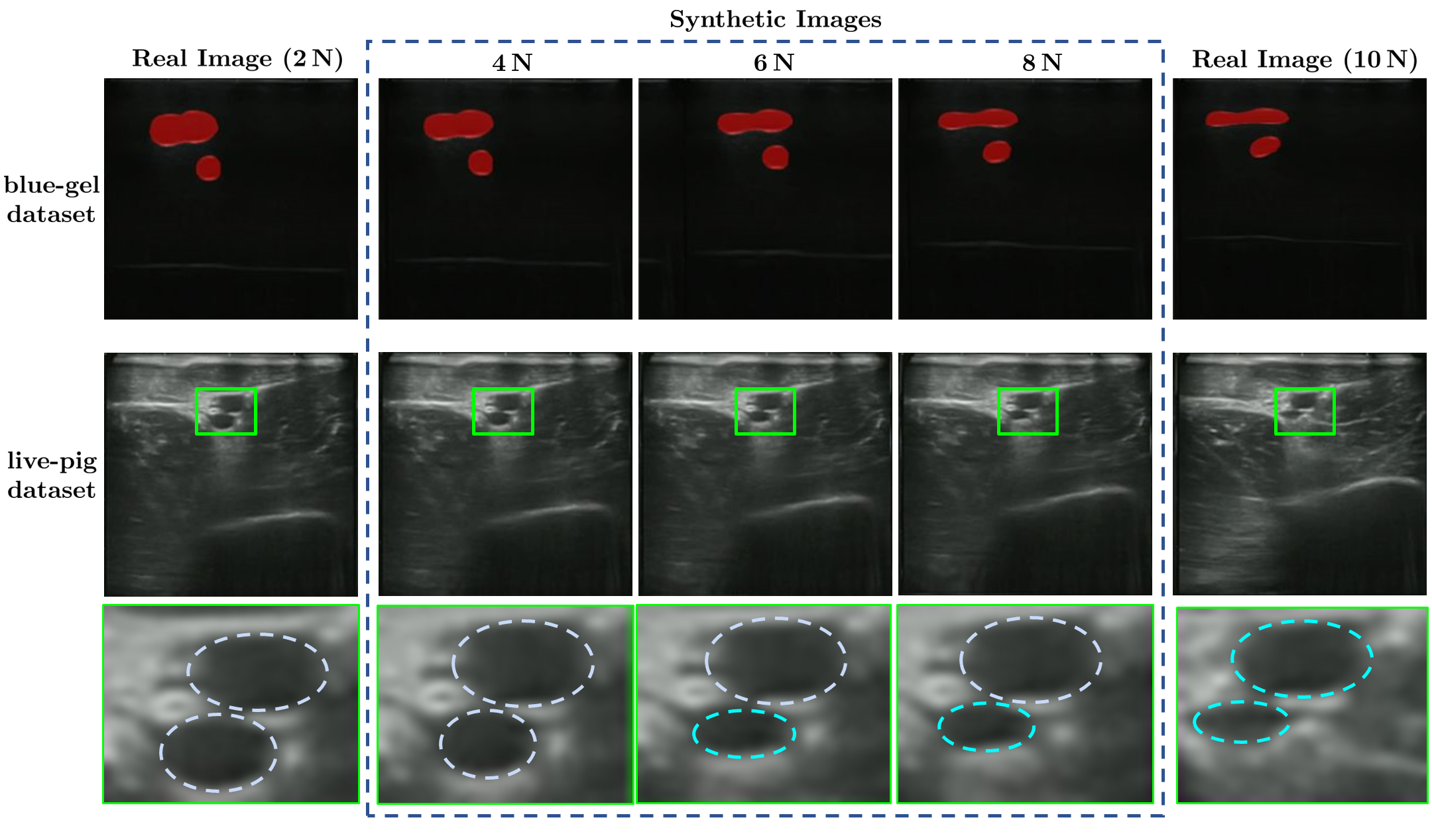}
    \end{center}
    \caption{Real and synthetic images from the blue-gel dataset and live-pig dataset are shown on the first and second row, respectively. The third row shows the zoomed-in view of vessels from the second row. From left to right, the images show a gradual compression of the vessel shape as the force is increased.}
    \label{fig:fig-synthetic-data}
\end{figure*}
\begin{figure} [ht]
    \begin{center}
        \includegraphics[width=\linewidth]{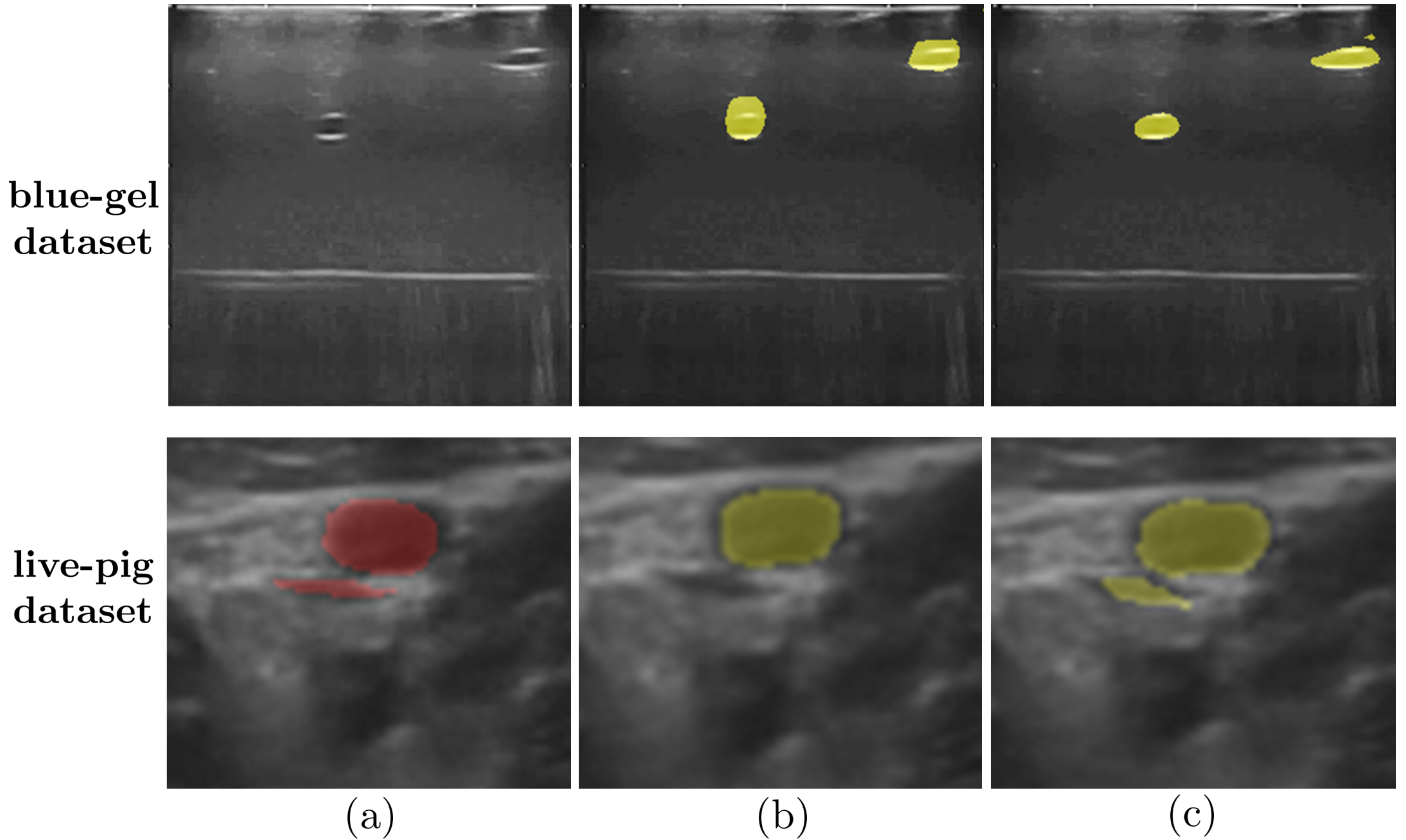}  
    \end{center}
    \caption{(a) Ultrasound image at 10\,N. (b) Ultrasound image with predicted segmentation mask using a U-Net model trained using real data collected at 2\,N. (c) Ultrasound image with predicted segmentation mask using U-Net model train using real and synthetic data. The inclusion of synthetic data improves segmentation accuracy. We also show a zoomed-in view of the live-pig dataset where we are able to segment an almost collapsed vein with the augmentation of the synthetic dataset.} 
    \label{fig:segmentation_viz}
\end{figure}
\subsection{Synthetic data generation for training deformed vessel segmentation models}
In this set of experiments, we evaluate the effect of using synthetic images to improve the results of vessel segmentation under tissue deformations. In the first experiment, we evaluated the realism of the synthetic ultrasound images created using the method described in Section \ref{sec:approach}-C on the blue-gel and live-pig datasets. Using the scanning mode of the robot, we collected $\sim$775 images on the blue-gel phantom, and $\sim70$ images on a live pig at 2\,N and 10\,N each. We then created synthetic images at 3 intermediate force values $F=4$\,N, $F=6$\,N, and $F=8$\,N, denoted as $D_{4,\textrm{syn}}$, $D_{6,\textrm{syn}}$, and $D_{8,\textrm{syn}}$, respectively. Example synthetic images are shown in Fig. \ref{fig:fig-synthetic-data}. We use SSIM to measure the similarity between $D_{4,\textrm{syn}}$, $D_{6,\textrm{syn}}$, $D_{8,\textrm{syn}}$ and $D_{4,\textrm{real}}$, $D_{6,\textrm{real}}$ and $D_{8,\textrm{real}}$, respectively. To show that our method does not introduce excessive additional noise into the generated images, we also evaluate the peak-signal-to-noise ratio (PSNR)\cite{sara2019image} to compare the noise-to-signal ratio between $D_{\textrm{syn}}$ and $D_{\textrm{real}}$, expressed in decibels (dB). Note that the maximum possible PSNR value for $8$-bit images is $\sim\,$48 dB. As shown in the results in Table \ref{Synthetic image comparision}, we see high SSIM scores as well as high PSNR for both datasets indicating high realism of the synthetic images. We observe slightly lower SSIM and PSNR scores for the live-pig dataset compared to the blue-gel dataset because of noisy \emph{in-vivo} ultrasound images and movement of the live animal during data collection.
\begin{table}[ht]
\begin{center}
\caption{SSIM and PSNR Between Real Images and  Synthetic Images at Interpolated Force Values}
\label{Synthetic image comparision}
\begin{tabular}{|c|c|c|c|c|}
\hline
\textbf{} & \multicolumn{2}{c|}{\textbf{blue-gel}} & \multicolumn{2}{c|}{\textbf{live-pig}} \\ \hline
\multicolumn{1}{|l|}{} & \multicolumn{1}{l|}{\textbf{SSIM}} & \multicolumn{1}{l|}{\textbf{PSNR}} & \textbf{SSIM} & \textbf{PSNR} \\ \hline
F=4N &0.98 &42.7  &0.81 &25.6 \\ \hline
F=6N &0.97 &40.8  &0.81 &25.6 \\ \hline
F=8N &0.97 &38.8  &0.78 &23.7 \\ \hline
\end{tabular}
\end{center}
\end{table}

We now show how the synthetic images generated by U-RAFT can be used as a data augmentation technique to improve a vessel segmentation model with respect to intersection-over-union (IoU). We use the $\sim\,$775 blue-gel images from Section \ref{sec:results}-B as our training dataset. We collected an additional $\sim\,$175 images each for testing and validation datasets. Regarding the live-pig dataset, we have obtained data from two porcine models. Our training dataset consists of the same $\sim\,$70 images per force value as in Section \ref{sec:results}-B, with an additional $\sim\,$260 validation and $\sim\,$300 testing images collected using the palpation mode of the robot. We denote our training datasets as $D_{2,\textrm{real}}^{\textrm{train}}$, $D_{4,\textrm{real}}^{\textrm{train}}$, $D_{6,\textrm{real}}^{\textrm{train}}$, $D_{8,\textrm{real}}^{\textrm{train}}$,$D_{10,\textrm{real}}^{\textrm{train}}$. 
The live-pig dataset has ten times fewer images than the blue-gel dataset because of high cost and effort involved in conducting \emph{in-vivo} porcine experiments. The vessel labels for these images are manually annotated by trained annotators with at least $6$ months of experience in labeling vessels in ultrasound images.

For this experiment, we train individual U-Net models on $D_{2,\textrm{real}}^{\textrm{train}}$, $D_{10,\textrm{real}}^{\textrm{train}}$ and a combination of $D_{2,\textrm{real}}^{\textrm{train}}+D_{10,\textrm{real}}^{\textrm{train}}$ dataset. Further, we augment the $D_{2,\textrm{real}}^{\textrm{train}}+D_{10,\textrm{real}}^{\textrm{train}}$ dataset with multiple-force synthetic data ($D_{4,\textrm{syn}}$, $D_{6,\textrm{syn}}$, and $D_{8,\textrm{syn}}$) and use it to train a U-Net model. We also compare our augmentation technique to the random elastic augmentation mentioned in \cite{ronneberger2015u} to underscore the significance of augmenting using multiple-force synthetic data. We show the improvement in segmentation on both the blue-gel and live-pig dataset. 
\begin{table}[ht]
\begin{center}
\caption{Comparing performance on the test datasets of a U-Net model with and without data augmention using the synthetic dataset. The results from the best performing dataset are highlighted.}
\label{tab:unet_results}
    \begin{tabular}{|c|c|ccc|}
\hline
 & \textbf{blue-gel} & \multicolumn{3}{c|}{\textbf{live-pig}} \\ \hline
\textbf{\begin{tabular}[c]{@{}c@{}}Training\\ Dataset\end{tabular}} & \textbf{\begin{tabular}[c]{@{}c@{}}IoU \\\end{tabular}} & \multicolumn{1}{c|}{\textbf{\begin{tabular}[c]{@{}c@{}}IoU\\ All Vessels\end{tabular}}} & \multicolumn{1}{c|}{\textbf{\begin{tabular}[c]{@{}c@{}}IoU\\ Arteries\end{tabular}}} & \textbf{\begin{tabular}[c]{@{}c@{}}IoU\\Veins\end{tabular}} \\ \hline
 $D_{2,\textrm{real}}^{\textrm{train}}$ & 0.77 & \multicolumn{1}{c|}{0.53} & \multicolumn{1}{c|}{0.58} & 0.48 \\ \hline
 $D_{10,\textrm{real}}^{\textrm{train}}$& 0.73 & \multicolumn{1}{c|}{0.49} & \multicolumn{1}{c|}{0.54} & 0.45 \\ \hline
 $D_{2,10,\textrm{real}}^{\textrm{train}}$ & 0.86 & \multicolumn{1}{c|}{0.56} & \multicolumn{1}{c|}{0.59} & 0.51 \\ \hline
 $D_{2,10,\textrm{real}}^{\textrm{train}}$+$D_{4,6,8,\textrm{syn}}^{\textrm{train}}$ & \textbf{0.89} & \multicolumn{1}{c|}{\textbf{0.62}} & \multicolumn{1}{c|}{\textbf{0.64}} & \textbf{0.58} \\ \hline
 $D_{2,10,\textrm{real}}^{\textrm{train}}$+$D_{\textrm{rand-syn}}^{\textrm{train}}$& 0.86 & \multicolumn{1}{c|}{0.56} & \multicolumn{1}{c|}{0.60} & 0.52 \\ \hline
\end{tabular}
\end{center}
\end{table}

As shown in Table \ref{tab:unet_results}, augmenting the real dataset with multi-force synthetic data outperforms the training done using only real images. We also see it outperform the model trained with random elastic deformations, highlighting the need for realistic force-based augmentation using deformable registration. We also compare the affect of force-based augmentation on veins and arteries separately in Table \ref{tab:unet_results}, as arteries and veins experience different levels of deformation, i.e. veins collapse easier than arteries. We see that there is a larger improvement in IoU for vessels that experience larger deformations, demonstrating that data augmentation with deformable registration is of particular importance when deformations are large. An example of this is shown in Fig. \ref{fig:segmentation_viz}, where a model trained using only on 2\,N images fails to segment a nearly collapsed vein, but the model trained with our synthetic images is able to segment the vein. 

\section{Conclusions and Future Work} \label{sec:conclusions}
In conclusion, this paper proposes an unsupervised approach for training an ultrasound-to-ultrasound deformable registration model. We propose and compare three different loss functions and show that a loss function based on feature-aware cyclic loss performs best. We also demonstrate how our approach, combined with a force-controlled robot, can be used to generate synthetic deformed images to expand the size of a femoral vessel segmentation training dataset and improve vessel segmentation performance. We also validate our approach on both a benchtop human tissue/vessel phantom and \emph{in-vivo} porcine images, highlighting the practical application of our deformable registration model in real-world medical imaging tasks. Overall, this paper presents an innovative solution to address the challenges associated with ultrasound imaging, particularly in the areas of image registration and segmentation which are critical for accurate vascular access.

As a part of future work, we plan to extend the U-RAFT algorithm with a recurrent model to learn a physics model in order to improve the deformation prediction especially for predicting scenarios like vessel decollapsing. To improve the live-pig IoU score, we will also like to further expand the training dataset by using the spatial augmentation  mentioned in \cite{morales2023reslicing}, along with force-based augmentation.

\section*{Acknowledgment}
We would like to thank Nico Zevallos, Dr. Michael R. Pinsky, Dr. Hernando Gomez, Lisa Gordon, Antonio Gumucio, Ted Lagattuta, and Andrew Schoenling, for their help in dataset collection and labelling. We would also like to thank Yizhu Gu for help with the design and fabrication of the system components.

\bibliographystyle{IEEEtran}
\bibliography{references}

\end{document}